# Universality, the Barton, Namikawa, and Nakajima relation, and scaling for dispersive ionic materials

*J. Ross Macdonald*
*Department of Physics and Astronomy, University of North Carolina,*
*Chapel Hill, NC 27599-3255, USA;, Tel.: +1-919-967-5005; macd@email.unc.edu*

Many frequency-response analyses of experimental data for homogeneous glasses and single-crystals involving mobile ions of a single type indicate that estimates of the stretched-exponential $\beta_1$ shape parameter of the Kohlrausch K1 fitting model are close to 1/3 and are virtually independent of both temperature and ionic concentration. This model, which usually yields better fits than others, is indirectly associated with temporal-domain stretched-exponential response having the same $\beta_1$ parameter value. Here it is shown that for the above conditions several different analyses yield the important and unique value of exactly 1/3 for the $\beta_1$ of the K1 model. It is therefore appropriate to fix the $\beta_1$ parameter of this model at the constant value of 1/3, then defined as the U model. It fits data sets exhibiting conductive-system dispersion that vary with both temperature and concentration just as well as those with $\beta_1$ free to vary, and it leads to a correspondingly universal value of the Barton-Nakajima-Namikawa (BNN) parameter $p$ of 1.65. Composite-model complex-nonlinear-least-squares fitting, including the dispersive U-model, the effects of the bulk dipolar-electronic dielectric constant, $\varepsilon_{D\infty}$, and of electrode polarization when significant, also leads to estimates of two hopping parameters that yield optimum scaling of experimental data that involve temperature and concentration variation.

I. INTRODUCTION

In 1994 Phillips suggested that the most important unsolved problem in physics today is to explain relaxation in complex (disordered) systems by microscopic analysis [1]. Roling and Martiny have recently stated that "Finding an explanation for this high degree of universality (i.e., the existence of master-curves for conductivity isotherms) is still one of the major challenges of solid state physics" [2]. The present work addresses both these challenges. The apparent universality of conduction in disordered solids has been discussed by Dyre and Schrøder who suggested various macroscopic and microscopic models that predict such universality in the extreme disorder limit [3]. Since universality is an idealized concept, it is not surprising that most claims for universal behavior and models [e.g., 4,5] have been found too limited or even incorrect, but this should not discourage new universality proposals, ones that are generally doomed eventually to suffer the same fate as new experimental results are analyzed and the domain to which the model is applied is sufficiently extended.

The principal aim of the present work is to describe a new universal, conductive-system frequency response expression, the U model; define its range of application; and demonstrate its usefulness in fitting and analyzing experimental immittance data. This model is simpler than others used in the past; it is derived from both macroscopic and microscopic analyses; and it and its three parameters are strongly physically based. Complex nonlinear least squares (CNLS) fitting of dispersive data using the U model allows highly accurate discrimination between bulk ionic response and that associated with electrode effects [6-10, and the present work]. Although such effects are usually important in the low-frequency region of experimental data, they may sometimes be non-negligible in the high-frequency region as well [8].

In Section II-A two general expressions for dispersion associated with charge carrier motion are described, while in II-B they are specialized by setting a general ion-ion correlation function involved in them to stretched-exponential temporal response, leading to the specific dispersive frequency-response K0 model and its superior transformed version, the K1 model. This well-fitting model is then simplified to yield the U model, and justifications for the fixed value of 1/3 of its shape parameter provided. Its limited universality and its prediction of a universal value for the BNN parameter are then discussed. In Section III, the U model is used for fitting and scaling of data for the same material at different temperatures and at different ionic concentrations, as well as demonstrating the separation of bulk dispersion and electrode effects. Finally, Section IV summarizes the conclusions of the work.

## II. DETAILS OF SOME CONDUCTIVE-SYSTEM DISPERSIVE MODELS

### A. General expressions

In order to distinguish easily between various fitting models, let the index $k$ take on values of 0 or 1, used in-line or as a subscript. Then for $k = 0$, if $\phi_0(t)$ is a conductive-system correlation function defined in the time domain, the corresponding normalized frequency response complex function, defined at the complex resistivity level, is [9,11]

$$I_0(\omega) = I_0'(\omega) - iI_0''(\omega) = \frac{\rho_{C0}(\omega) - \rho_{C0}(\infty)}{\rho_{C0}(0) - \rho_{C0}(\infty)} = \int_0^\infty \exp(-i\omega t)\left(-\frac{d\phi_0(t)}{dt}\right)dt, \qquad (1)$$

involving a one-sided Fourier transform. Note that a specific expression for the $I_0(\omega)$ response model only follows when one is specified for $\phi_0(t)$, as in the next section. When the small or zero $\rho_{C0\infty} \equiv \rho_{C0}(\infty)$ quantity is neglected as usual [12], the corresponding $k = 0$ frequency response at the complex modulus level is just $M_{C0}(\omega) = M_{C0}''(\omega) + iM_{C0}''(\omega) = i\omega\varepsilon_V\rho_0 I_0(\omega)$. Here the subscript "$C$", used for theoretical and model quantities, denotes conductive-system response, and $\varepsilon_V$ is the permittivity of vacuum. We shall not distinguish between the $k = 0$ and 1 dc resistive quantities $\rho_{C0}(0)$ and $\rho_{C1}(0)$, and will thus use $\rho_0 = 1/\sigma_0$ for either. The quantity $\sigma_0$ is the dc limit of the real part of the conductivity, $\sigma(\omega) = \sigma'(\omega) + i\sigma''(\omega) \equiv 1/\rho(\omega)$, where $\rho(\omega) = \rho'(\omega) - i\rho''(\omega)$. The corresponding complex dielectric constant expression is $\varepsilon(\omega) = \varepsilon'(\omega) - i\varepsilon''(\omega) \equiv 1/M(\omega)$.

Consider now the different $k = 1$ response, closely related to $I_0(\omega)$, as demonstrated in the equation below. When the famous 1973 continuous-time, random-walk (CTRW) approximate microscopic model of Scher and Lax [13] is extended slightly to make its imaginary part fully consistent with its real part at the complex conductivity level [12], this conductive-system hopping model may be expressed most simply at the complex modulus level as [9,12]

$$M_{C1}(\omega) = M'_{C1}(\omega) + iM''_{C1}(\omega) = i\omega\varepsilon_V\rho_0 I_1(\omega) \equiv [1 - I_{01}(\omega)]/\varepsilon_Z, \qquad (2)$$

where the important effective-dielectric-constant quantity $\varepsilon_Z$ is defined as $\varepsilon'_{C1}(\infty) \equiv \varepsilon_{C1\infty} = 1/M_{C1}(\infty)$, and the 01 subscript here indicates that $I_{01}(\omega)$ is of the form of $I_0(\omega)$ but it involves $I_1(\omega)$ fit parameters rather than those obtained by direct fitting of the same data with the $k = 0$ $I_0(\omega)$ model. In Ref. 13, the $\phi_0(t)$ correlation function that leads to $I_0(\omega)$ normalized frequency response is defined as the probability that a hopping entity remains fixed in place over the time interval from 0 to $t$.

Equation (2) provides a direct connection between the different $I_1(\omega)$ and $I_0(\omega)$ responses, but the time domain response following from $I_1(\omega)$ is not the same as that of $I_0(\omega)$ [12]. Importantly, Eq. (2), arising from a detailed microscopic analysis, is of exactly the same form as that derived macroscopically, contemporaneously, and independently by Moynihan, Boesch, and Laberge [14] by considering electric field decay at constant dielectric displacement. This formal micro-macro agreement, unique for this $k=1$ model, thus provides additional justification for it. It remains general, however, until specific forms of $\phi_0(t)$ and $\varepsilon_Z$ are introduced.

It follows from the work of Scher and Lax [13], Eq. (2), and Refs. [6-9,12, and 15] that for the general present $k=1$ dispersion model the very important quantity $\varepsilon_Z = \varepsilon_{C1\infty}$ may be expressed as

$$\varepsilon_{C1\infty} = (\sigma_0/\varepsilon_V)/<\tau^{-1}>_1 = \varepsilon_{Ma}<x>_{01} = [\gamma N(qd)^2/(6k_B\varepsilon_V)]/T, \qquad (3)$$

a purely conductive-system quantity. Here $x \equiv \tau/\tau_o$; $\tau_o$ is a characteristic relaxation time for the model that determines the placement on the frequency scale of the model response; $\varepsilon_{Ma} \equiv \sigma_0\tau_o/\varepsilon_V$; and $<\tau^{-1}>_1$ and $<\tau>_{01} \equiv \tau_0<x>_{01}$ are different averages over the $k=1$ and $k=0$ distributions of relaxation times, respectively. When the form of $\phi_0(t)$ is known, one can calculate $<\tau>_0$ from $<\tau>_0 = \int_0^\infty \phi_0(t)\,dt$ [11,13]. Equation (3) is consistent with the Scher-Lax result that $\rho_0$ is proportional to $<\tau>_{01}$, identified in their work as the mean waiting time for a typical hop [12,13,15], a physically plausible result.

The quantity $N$ is the maximum mobile charge number density; $\gamma$ is the fraction of charge carriers of charge $q$ that are mobile; $d$ is the rms single-hop distance for a hopping entity, and $k_B$ is the Boltzmann constant. The high-frequency-limiting effective dielectric constant, $\varepsilon_{C1\infty}$, associated entirely with mobile-charge effects, is likely to arise from the short-range vibrational and librational motion of caged ions. In addition to $\varepsilon_{C1\infty}$, a bulk high-frequency-limiting dielectric constant, $\varepsilon_{D\infty}$, associated with non-dispersive dipolar and vibratory effects, is always present. Thus, for an appreciable range of high-frequencies [12], the total limiting dielectric constant is $\varepsilon_\infty = \varepsilon_{C1\infty} + \varepsilon_{D\infty}$. Although the important quantity $\sigma'(\omega)$ approaches a final plateau at sufficiently high frequencies [12], we shall be concerned here only with high-frequency-limiting behavior occurring before the effects of the plateau become important, the usual experimental situation.

Because of the endemic presence of $\varepsilon_{D\infty}$, a quantity not directly associated with mobile-charge effects, it is necessary in fitting data to always include a free dielectric parameter in the fitting model, $\varepsilon_x$, to represent $\varepsilon_\infty$ for $k=0$ or $\varepsilon_{D\infty}$ for $k=1$ situations, as discussed in the following section. Two fitting parameters present in all the following models are $\rho_0$, a quantity that determines the magnitude or scale of the response, and $\tau_o$, defined above.

### B. Specific models associated with stretched-exponential temporal response: K0, K1, and U

*1. Stretched-exponential temporal response, the K0 frequency-response model, and electrode effects*

As already mentioned, an explicit expression for $\phi_0(t)$ is required in order to calculate specific $k=0$ and $k=1$ frequency responses for data fitting or data simulation. The $k=0$ choice leads to K0-model response when the ubiquitous stretched-exponential relation, $\phi_0(t) = \exp\{-(t/\tau_o)^{\beta_0}\}$ with $0 < \beta_0 \leq 1$ [1,16,17], originally introduced by Kohlrausch, is used in Eq. (1) to obtain an explicit expression, or numerical representation, for $I_0(\omega)$. Then, such results may be used to calculate the K0-model modulus-level frequency response $M_{C0}(\omega) = i\omega\varepsilon_V\rho_0 I_0(\omega)$ [6-10]. Next, Eq. (2) leads to K1-model frequency response [7,9,18]. Here $\beta_0$ is both the stretching factor in the time domain as well as the parameter that determines the shape of K0-model frequency response. The corresponding K1 shape parameter, unequal to $\beta_0$, is identified as $\beta_1$. Although both the K0 and K1 model responses must actually be calculated numerically for arbitrary $\beta_k$ values, the free LEVM CNLS computer program allows the parameters of these models to be very accurately determined for both data fitting and simulation tasks [19].

Unlike the K1 model, which involves the non-zero high-frequency-limiting effective dielectric constant $\varepsilon_{C1\infty}$ of Eq. (3), conductive-system K0 response involves no such limiting value and so $\varepsilon'_{C0}(\infty) \equiv \varepsilon_{C10} = 0$ [6,9]. But for experimental data that are well fitted by the K1 model, the actual total high-frequency-limiting dielectric constant implicit in the data is $\varepsilon_\infty = \varepsilon_{C1\infty} + \varepsilon_{D\infty}$, and it will be this value, rather than just $\varepsilon_{D\infty}$, that is estimated by the $\varepsilon_x$ free parameter that must be included in fitting using the K0 model. Such a composite model has been designated the CK0, where the "C" here represents the capacitance associated with $\varepsilon_x$ [7,9].

For most data situations, one should also include in a composite fitting model a separate electrode-effects model in series with that representing bulk conductive-system dispersion. The electrode model should represent the effect of partial or complete blocking of mobile charges at the electrodes. Surprisingly, it has been shown that such effects can sometimes be important at high frequencies, as well as at the low-frequency end of the data range [8]. Since electrode effects may thus appreciably influence experimental data, as demonstrated in Section III-B below as well as in Ref. 8, it is important to initially include this possibility in a composite fitting model and evaluate the need for such inclusion.

In recent work [9], CNLS fitting results, using several different Kohlrausch and other fitting models, with and without electrode-effect contributions, have been compared using experimental data for a glass in single-crystal form. The inclusion of electrode effects led to important improvements in fit accuracy for most of the model fits, particularly for the best-fit K0 and K1 ones. Even when not mentioned explicitly in the following work, it should be understood that the fits of experimental data sets discussed herein included not only a bulk-dispersion model such as the K1 or U one, but also a series electrode-model contribution.

## 2. *Two different K1 models*

a. The original-modulus-formalism fitting model

The widely used pioneering treatment of Ref. [14], now termed the original modulus formalism (OMF) approach and involving the $k=1$ K1 response model defined in Eq. (2), is unfortunately critically flawed by its improper identification of the $\varepsilon_Z$ of Eq. (2) as $\varepsilon_\infty \equiv \varepsilon'(\infty)$, a quantity that includes all contributions to the high-frequency-limiting dielectric constant [18]. Since the authors did not recognize the existence of $\varepsilon_{C1\infty}$, their $\varepsilon_\infty$ was considered to be just $\varepsilon_{D\infty}$, rather than $\varepsilon_\infty = \varepsilon_{C1\infty} + \varepsilon_{D\infty}$. In the usual case where both quantities are non-zero, data fitting would yield an estimate of $\varepsilon_x = \varepsilon_\infty$, identified as $\varepsilon_{D\infty}$, but actually including both contributions to $\varepsilon_\infty$ [7,9,18]. The failure to distinguish between these two quantities by not including a separate fit parameter such as $\varepsilon_x$ leads to an inappropriate mixing of dipolar dielectric effects and those associated only with mobile charge, and thus to both theoretical and experimental inconsistencies [7,9,18].

Many hundreds of published data fits and analyses since 1973 of $M''(\omega)$ data using the OMF, and thus the K1 model alone, have yielded strong dependence of the estimated $\beta_1$ values on ionic concentration and appreciable temperature dependence as well. For example, as the ionic concentration approaches zero, OMF fits lead to $\beta_1$ estimates that approach unity [e.g., 20]. This is because then $\varepsilon_{C1\infty} \to 0$, $\varepsilon_\infty \to \varepsilon_{D\infty}$, and true dispersive effects become more and more negligible compared to Debye-type relaxation involving only $\sigma_0$ and $\varepsilon_{D\infty}$, response that necessarily involves a $\beta$ value of unity. It is clear that all OMF fits should be fully discounted, and such fitting replaced by a consistent approach such as the corrected modulus one described below.

b. The corrected-modulus-formalism approach

The corrected modulus formalism (CMF) also uses the K1 model but includes a free $\varepsilon_x$ parameter, therefore denoted the CK1 model. For this model, $\varepsilon_x = \varepsilon_{D\infty}$ because $\varepsilon_{C1\infty}$ is not a free parameter of the fit and is completely determined, as in Eq. (3), by the estimates of the K1-model free parameters: $\sigma_0$, $\tau_o$, and $\beta_1$. It has been found that CK1 fits for a variety of materials, ionic concentrations, and temperatures lead to virtually constant estimates of $\beta_1$, all very close to a value of 1/3, along with both better fitting and no inconsistencies [6-9,18,21].

Let us temporarily replace the symbol $\beta_1$ by $\beta_{1C}$ to distinguish it from a $\beta_1$ obtained from OMF fitting. Then for the K1 model with $\varepsilon_Z = \varepsilon_{C1\infty}$ in Eq. (2), Eq. (3) may be expressed as

$$\varepsilon_{C1\infty} = \varepsilon_{Ma} <x>_{01} = \varepsilon_{Ma} \beta_{1C}^{-1} \Gamma(\beta_{1C}^{-1}) = A/T , \qquad (4)$$

appropriate for CK1 fits. Here $A$ is the term in square brackets at the right-side end of Eq. (3) and depends on ionic concentration but not appreciably on temperature [7]; $\Gamma(\ )$ is the Euler gamma function; and $\beta_{1C}$ is a value of $\beta_1$ obtained from fitting using the CMF with the separate free parameter $\varepsilon_x$ to estimate $\varepsilon_{D\infty}$. It follows from Eq. (4) that when $A$ is temperature independent, the thermally activated quantities $T\sigma_0$ and $\tau_o$ each exhibit Arrhenius behavior and their product is itself temperature independent.

3. *The U model and some of its consequences*

a. $\beta_1$ derivations

The U model, a simplification of the corrected modulus formalism approach, is particularly important both because of its simplicity (only two free parameters) and because of its universal character over a wide (but still limited) domain of applicability. It is defined as a Kohlrausch K1 model in which the important shape parameter, $\beta_1$, is fixed at the value of 1/3. In practice, it will represent the conductive-system dispersive-model part of a composite fitting model that includes not only $\varepsilon_{D\infty}$ but also a part that accounts for electrode effects when important.

The criteria that define the extent of the universality of the U model are that it applies only for homogeneous materials that potentially allow conduction in all three dimensions and involve mobile charge carriers of a single type [21]. Only materials and data satisfying these conditions are discussed here for U-model applications. Fitting results with the CK1 model and $\beta_1$ taken free to vary may be expected not to satisfy one or more of these criteria when the estimate of $\beta_1$ is appreciably different from 1/3.

In the past, data fitting with either the original or corrected modulus formalism, both of which involve the K1 model, has involved a $\beta_1$ parameter usually taken free to vary and therefore determined by the fit estimate. For such analyses, either no physical explanation of $\beta_1$ values has been suggested or they have been improperly interpreted, on the basis of inappropriate original-modulus-formalism fit results, as a measure of correlation between hopping ions [*e.g.*, 22]. It is therefore of great importance to provide experimentally and physically based justifications for the fixed value of 1/3, one that cannot be interpreted as associated with variable ion-ion correlation. Three different approaches are presented below.

i. <u>Experimental.</u>

Define $n$, $n_0$, and $n_1$ as the high-frequency-limiting log-log slopes of $\sigma'(\omega)$ for data, and for the K0 and K1 models, respectively. Data fitting and analysis show that these quantities are closely frequency-independent for sufficiently high frequencies in the absence of nearly constant loss and high-frequency electrode effects, and so they are the exponents of power-law responses and fitting with a power-law model is appropriate for determining $n$. Good fits of $\sigma'(\omega)$ experimental and synthetic data extending to high-frequencies indeed show that all three of these slopes are equal as they should be, and for materials satisfying the present U-model criteria $n$ is frequently found to be very close to 2/3 in value for many different glasses [23-25].

Fitting estimates of $\beta_0$ and $\beta_1$ using the CK0 and CK1 models, lead to $\beta_0 = n_0 = n$ and to $1 - \beta_1 = n_1 = n$. The usual relation [9,12] $1 - \beta_1 = \beta_0$ follows immediately. When the limiting slopes are 2/3, it follows that $\beta_0 = 2/3$ and $\beta_1 = 1/3$, the value used in the U model. Therefore, it seems likely that the U model would be most appropriate for fitting all data for which $n \simeq 2/3$.

ii. <u>Hopping theory.</u>

Although Phillips [1,26] has treated stretched exponential relaxation in great detail, his work primarily considered mechanical and dielectric relaxation results for non-conductors, with little consideration of $\beta_1$ estimates obtained from data involving frequency dispersion associated with mobile charge carriers. As he discusses, however, several treatments of the trapping model involving fixed trapping sites lead to the result

$$\beta = d_e / (2 + d_e), \tag{6}$$

where $d_e$ is the effective dimensionality of the configuration space in which dispersive effects occur.

For Coulomb interaction, the value $d_e = 3/2$ was derived, leading to $\beta = 4/7$, while for spin glasses on cubic lattices a value of 1/3 was obtained with $d_e = 1$. The spin-glass model is not directly appropriate for the present situation, and the maximum-disorder universal-model treatments of Dyre and Schrøder [3] do not involve explicit Coulomb effects. A value of $d_e = 1.35$ was found to be most appropriate by these authors. Further, there is no reason to believe that the stretched-exponential correlation function, $\phi_0(t)$ with $0 < \beta_0 \leq 1$, necessarily includes such effects.

The U model is consistent with a value of $d_e = 1$ when Eq. (6) is applicable. Some implications of the above results are as follows. The Scher-Lax stochastic model, the microscopic basis for the K1 and U models [13], is a 3D one that treats all sites on a discrete lattice as equivalent and independent and leads to frequency response substantially different from that of a later on-off one-dimensional bond-percolation stochastic model [27,28]. Thus, the $\beta_1 = 1/3$ value need not directly imply that the motion of the hopping charge carriers is one dimensional, but it implies that the stretched-exponential correlation function determining the K1 response is associated with a waiting-time distribution best interpreted in terms of correlated processes occurring in a configuration space with an effective dimension of unity. In contrast, if Eq. (6) were applied directly to the K0 model, then the $\beta_0 = 2/3$ value would be associated with an implausible effective dimensionality of 4. These conclusions raise the need for a detailed microscopic treatment that justifies the U-model requirement that the effective dimensionality of its ion-ion correlation function be unity.

Although a recent geometric derivation of stretched-exponential response for mobile electrons leads to a continuously variable $\beta$ [29] and so is not relevant to the present results, a much earlier CTRW treatment [16,17], involving some elements of the Scher-Lax Eq. (2)-model derivation, showed that the value of $\beta$ in stretched-exponential response was determined by the rate at which a hopping entity finds a new site. Thus if $\beta(1)$ is the value of $\beta$ for one-dimensional hopping, then for motion in three dimensions, $d = 3$ and $\beta(3) = 2\beta(1)$. If we identify $\beta_0$ as $\beta(3)$ and $\beta_1$ as $\beta(1)$, as indicated by U-model fitting results, then we may write $\beta(3) = 1 - \beta(1) = 2\beta(1)$, whose solution is $\beta(1) = 1/3$. Thus the above two relations require not only that $\beta(1) = 1/3$ and $\beta(3) = 2/3$ but they show that these are the *only* values consistent with $\beta(3) = 2\beta(1)$.

iii. Topological and conclusions

A pending treatment of the motion of ions of a single type in homogeneous materials makes use of physically based topological considerations [21], not to be confused with geometrical ones. The analysis starts with the recognition that a forcing electric field present between two charged plane-parallel electrodes induces a uniaxially anisotropic local dynamical metric. Within a local polar coordinate frame there is a radial coordinate and $(d-1)$ angular coordinates. For the $d = 3$ situation, local motion with respect to the azimuthal coordinate is irrelevant for homogeneous materials, so the effective dimensionality is $d_e = 2$, while for streaming motion transverse to the electrodes $d_e = 1$. When the approach is applied to temporal stretched-exponential behavior, it leads to just $\beta = d_e/d$, consistent with the present results if $d_e = 2$ for the K0 model and $d_e = 1$ for the K1 one.

A natural interpretation is that for high-frequencies, where hopping motion is local, both models should lead to a limiting slope of 2/3, as observed for both synthetic and experimental data. The motion of the charges at very low frequencies should be of streaming one-dimensional character, consistent with the K1 correlation function involving $\beta_1 = 1/3$ and with the observation that synthetic K1-model response transformed to the time domain is not of stretched-exponential character except in the limit of long times where its stretching parameter is indeed 1/3.

The three disparate approaches above all lead to the same unique value of $\beta_1 = 1/3$ for the U-model and to the corresponding K0 value of $\beta_0 = 2/3$. Although both models involve the same high-frequency limiting slope of 2/3, their responses are different except in the extreme high-frequency region, and one generally finds that U-model fits of appropriate experimental data are much better than K0 ones with $\beta_0$ fixed at 2/3 or free to vary.

b. Consequences of the $\beta_1 = 1/3$ requirement and the BNN relation

Now when one sets $\beta_1$ fixed at 1/3 in the K1 model of Eqs. (2) and (4) to obtain the U response model, one finds that $\varepsilon_{C1\infty} = 6\varepsilon_{Ma}$, $\varepsilon_{C10} = 60\varepsilon_{Ma}$, and so $\Delta\varepsilon_{C1} \equiv \varepsilon_{C10} - \varepsilon_{C1\infty} = 54\varepsilon_{Ma} = \Delta\varepsilon$, since $\varepsilon_{D\infty}$ subtracts out from the experimental $\Delta\varepsilon \equiv \varepsilon_0 - \varepsilon_\infty$, one of the virtues of using $\Delta\varepsilon$ rather than either of its two components. In addition, if one defines $\varepsilon''_{C1S}(\omega) \equiv \varepsilon''_{C1}(\omega) - (\sigma_0/\omega\varepsilon_V)$, then the resulting dielectric loss arising from charge motion rather than from dipolar dispersion involves a peak response at $\nu_p \equiv \omega_p/2\pi \cong 0.01122/2\pi\tau_o$, with $\varepsilon''_{C1S}(\omega_p) \cong 14.405\,\varepsilon_{Ma}$ for the peak value.

An empirical expression that has been of considerable importance in the past is that of Barton [30], Nakjima [31], and Namikawa [32], commonly known as the BNN relation,

$$\sigma_0 = p\varepsilon_V \Delta\varepsilon \omega_p, \qquad (5)$$

where $p$ is a numerical constant of order 1, and $\omega_p$ is the radial frequency dielectric loss peak, only equal to the value listed above for the U model. For that model, however, it follows that $p = 1/(0.01122 \times 54) \cong 1.65$, a universal value for all data that are well fitted by the U model.

Figure 3 in Ref. 3 is a log-log BNN-related plot that indicates that most estimates of $p$ are close to the above value, a satisfying result in view of the usual uncertainties associated with estimates of $\Delta\varepsilon$ and $\omega_p$ from experimental data. Although Porto et al. [33] have recently questioned the applicability of the BNN relation under changes in charge carrier concentration, excellent U-model fittings for wide concentration variation [7], and the results discussed below, show that estimates of $p$ from such fits confirm the 1.65 value. Thus, it will usually be appropriate in future to replace the BNN relation by any of those listed above that connect effective dielectric quantities, such as $\varepsilon_{C1\infty}$, to $\varepsilon_{Ma}$.

### III. FITTING AND SCALING RESULTS FOR THE U MODEL

When one has available an excellent fitting model applicable for a particular experimental and material domain, there is no need for scaling since data fitting with such a model leads to explicit parameter estimates and thus to more information than does development of a master scaling model. Although the U model is thus superior to scaling within its domain of applicability, it is instructive to discuss scaling parameters following from it and to show their applicability for data that include variation of both temperature and relative ionic concentration, $x_c$. For this purpose, data for the following materials will be used, as listed in Table I: $0.5\text{Li} \bullet 0.5\text{La} \bullet \text{TiO}_3$ [34], $x_c\text{K}_2\text{O} \bullet (1-x_c)\text{GeO}_2$ [20], and $0.3(0.6\text{Na}_2\text{O} \bullet 0.4\text{Li}_2\text{O}) + 0.7\text{B}_2\text{O}_3$ [35].

## A. Scaling possibilities and limitations

Before presenting fitting and scaling results for these materials, it is desirable to consider scaling approaches. It has been customary to write a general scaling relation in the form

$$\sigma'(\omega)/\sigma_0 \equiv \rho_0 \sigma'(\omega) = F'(\omega \tau_S) \equiv F'(\nu/\nu_S), \qquad (7)$$

where the left-hand parts refer to data and the right-hand ones to a fitting model, and $\nu_S \equiv 1/(2\pi\tau_S)$ is the scaling frequency. The essence of good scaling then involves choosing appropriate $\nu_S$ scaling values. Usually, no fitting is actually carried out, and an equation such as (7) is merely written to define the type of scaling to be used for $\sigma'(\omega)$ data. Various explicit choices for $\nu_S$ and discussions of the historical background of scaling appear in Refs 2, 3, 6, and 33. Here, scaling will be carried out employing a fully complex version of Eq. (7) for fitting, one that may be used to fit complex data at any immittance level and may include non-hopping processes.

Of the many past choices for $\nu_S$, we here consider only those of Sidebottom [5], Dyre and Schrøder [3], Macdonald [6], and Roling and Martiny [2]. The first two are essentially equivalent, are both related to the BNN relation, and when expressed in terms of $\varepsilon_{Ma}$ are: $\nu_S \equiv (\varepsilon_{Ma}/\Delta\varepsilon)\tau_o^{-1}$ and $\nu_S \equiv (\varepsilon_{Ma}/\Delta\varepsilon)/(2\pi\tau_o)$, respectively. For the U model, where $\tau_S$ becomes the explicit characteristic response time of the model, $\tau_o$, estimated from data fitting, these results lead to $\nu_S = 0.0185/\tau_o$ and $\nu_S = 0.00295/\tau_o$, respectively. In contrast, the earlier work of the author [6] involved just the $\tau_o$ quantity derived from CMF fits of the K1 model. For the U model, that result becomes $\nu_S \equiv 1/(2\pi\tau_o) = 0.1592/\tau_o$. Finally, Roling and Martiny set $\nu_S \equiv \nu_p$, where $\nu_p$ is the frequency at the peak of the mobile-charge dielectric loss response, discussed in Section II-B. For the U model, the frequency of the $\varepsilon''_{C1S}(\omega_p)$ peak is $\nu_p = 0.001786/\tau_o$, a factor of exactly the U-model BNN $p$ value of 1.65 smaller than the Dyre and Schrøder result,

The differences in the numerical values of the numerators of the expressions for all of the above scaling quantities are not significant for scaling, so they are all equivalent in this sense. Nevertheless, the choice of a proper value of $\tau_o$ is of the greatest importance for good scaling. Note, however, that since $\Delta\varepsilon$ and $\nu_p$ may be directly estimated from experimental data without fitting, a good estimate of $\tau_o$ is not actually necessary to form a scaling value of $\nu_S$ when a value of $\sigma_0$ is available, also necessary for scaling of $\sigma'(\omega)$ itself.

But consider the following: the estimation of $\Delta\varepsilon$ from data requires separate estimated values of both $\varepsilon_0$ and $\varepsilon_\infty$. Estimation of both of these quantities, especially that of $\varepsilon_0$, is rendered uncertain by the usual presence of electrode effects, as demonstrated in the next section; in addition the data often do not extend to high enough frequencies to yield a good estimate of $\varepsilon_\infty$. Further, estimation of $\nu_S \equiv \nu_p$ requires subtraction using a good estimate of $\sigma_0$ and then the determination of the frequency of the peak of a curve that usually varies slowly in the neighbourhood of the peak, again an inherently inaccurate process for ordinary experimental data.

### B. U-model fitting

The above discussion shows that the usual determination of scaling factors may depend on the use of only one or two points of the data, rendering results uncertain. On the other hand, CNLS estimation of values of $\sigma_0$, $\tau_o$, and $\varepsilon_{C1\infty}$ from U-model fitting makes use of all the data in an optimum way, also provides an estimate of $\varepsilon_{D\infty}$, and allows electrode effects to be adequately accounted for as part of the fitting. Therefore, scaling with values of these quantities so estimated is much more appropriate than are other approaches. Let us define the resulting scaled variables as $\bar{\sigma} \equiv \sigma/\sigma_0$ and $\bar{\nu} \equiv \omega\tau_o \equiv \bar{\omega} \equiv \nu/\nu_S$, where $\nu_S \equiv 1/(2\pi\tau_o)$.

Fitting results are presented in Table I. All fits of experimental data shown here used the U model with an extra free parameter to estimate $\varepsilon_{D\infty}$ and usually with additional free parameters (not shown) to account for electrode polarization effects. Other fits of the present data using the K1 model with $\beta_1$ free to vary led to estimates of it that were, as usual, very close to 1/3. See also the fitting results of Refs. 7 and 21 for results for other materials. Even when the relative standard deviation, $S_F$, of one of these fits with $\beta_1$ free was slightly smaller than that obtained with it fixed at 1/3, the relative standard deviations of the free parameters were smaller than those with it free, indicating a more significant fit. It is also worth emphasizing that comparison of the $\tau_o$ estimates for the two types of fit show that they are far less stable than those of $\beta_1$ since a small change in the estimate of the latter results in an appreciable change in the corresponding $\tau_o$ estimate.

The results shown in the A and B and C and D rows in the table are consistent with earlier work where K1-model fits of the present kind led to $\beta_1 \cong 1/3$ estimates that were nearly independent of both temperature and ionic concentration, making it reasonable to use the fixed value of $\beta_1 = 1/3$ in the present work. It is worth noting, however, that the A and B row estimates of $\varepsilon_{C1\infty}$ are not exactly proportional here to $1/T$, as expected from Eq. (4) and from earlier work [7]. Although this discrepancy may be associated with the large role that electrode effects play in the present data, as indicated in the response curves presented below, it is somewhat more likely to be associated with non-Arrhenius behavior stemming from a physically reasonable low-end cutoff of the K1 distribution of relaxation times.

The U line in the table involves only K1-model response with $\beta_1$ fixed at a value of 1/3 and the other two parameters each having their scaled values of 1. When, in addition, $\varepsilon_V$ is also set to the scaled value of 1, then from Eq. (3) $\bar{\varepsilon}_{C1\infty} = 6$. All of the K0 results in the table involved real-part fitting except the listed $\bar{\varepsilon}_{C1\infty}$ value where full complex-data fitting was used. Such fitting necessarily used both the K0 model and a free dielectric-constant $\varepsilon_x$ parameter to estimate the $\bar{\varepsilon}_{C1\infty} = 6$ in the U-model scaled data. The scaled $\bar{\sigma}'(\bar{\nu})$ U-model data, designated UM, extended up to a maximum value of $10^5$, where the $n$ slope was about 0.664, and was fitted to the K0 model using nonlinear least-squares. The last column in the K0 row shows its $\beta_0$ estimate in square brackets. An appreciably larger frequency range would be needed for the resulting $\beta_0$ estimate to more closely approximate its limit of 2/3. For much data where $\bar{\sigma}'(\bar{\nu})$ is no larger than $10^2$ to $10^3$, CMF CK1-model fitting still leads to $\beta_1 \simeq 1/3$ estimates but CK0 fitting yields $0.5 < \beta_0 < 0.6$, even though power-law fitting of the part of the $\sigma'(\omega)$ data at the high frequency end usually results in $n$ estimates much closer to 2/3.

Note that although the present K0-model fit of the virtually exact K1 scaled synthetic data leads to a value of $\bar{\varepsilon}_{C1\infty}$ reasonably close to the exact value of 6, the estimate of $\bar{\tau}_o$ is about 18 rather than 1. Such a larger value than that for the K1 model is characteristic of K0 fits. Finally, the other quantities in the last column of the table are percentage $S_F$ values for fits of each set of A-D data after subtracting the effects of estimated $\varepsilon_{D\infty}$ and electrode polarization parameters from the original data and then fitting the subtracted results to the scaled U model. As expected, such subtraction of comparable quantities leads to less accurate fits than those involving the full data because it involves the subtraction of nearly equal large quantities to find their small differences. Nevertheless, the two parameter estimates of the model were very close to the exact U-model data values of unity for all four fits of subtracted data.

### C. U-model scaling

Complex-plane plots of resistivity data are particularly useful in showing low-frequency electrode effects when present. In order to compare curves for different materials and conditions, all the figures presented here involve data scaled as above using $\rho_0$ and $\tau_o$ values estimated from the unscaled U-model fits of Table I. Figure 1 presents such results for the A and B material listed in the table. In order to maximize clarity, not all points used in fitting are included in this and the other figures and in none does the size of a data symbol indicate its error bar. In Figure 1 only about half of the data points are plotted, and, in addition, for the B data the right-hand spur extending to over 1.3 for the actual fitting was cutoff as shown. No such cutoff was applied for the A results.

The figure shows that the original (scaled) data lines for the two temperatures are similar for the high-frequency response region but begin to diverge at the lowest frequencies. The remaining two curves are those for data from which the effects of $\varepsilon_{D\infty}$ and electrode polarization have been subtracted before scaling, a simple procedure after LEVM fitting. It is clear that in the present representation the contributions to the overall frequency response from non-hopping $\varepsilon_{D\infty}$ and series electrode-effects dominate the dispersive U-model hopping ones except at the high-frequency end of the curves. Further, we see that the remaining hopping points, identified by "sub" in the figure, fall closely on the U-model master-curve solid line, although those for the A situation show a bit more deviation than do the B ones.

It is particularly important to emphasize that, to the degree that the non-hopping effects were adequately estimated by the fit of the full data, the hopping response shown here does not consist of points fitted to the master curve but instead it represents the best *hopping-data* estimates obtained from the fits of the full data. The excellent agreement of the hopping points with the exact U-model master curve for both temperatures shows not only that the present scaling is appropriate but that the hopping response is indeed very well described by the U model and its restriction to $\beta_1 = 1/3$. As we shall see, these conclusions are further confirmed by the results shown in the subsequent figures. Fitting with a composite model to allow the non-hopping contributions to the full response to be subtracted does not guarantee that the resulting data points will lie close to the U-model hopping curve. That they actually do so is confirmation that the U model represents the dispersive hopping part of the response adequately.

Figure 2 shows a more stringent fitting situation; one where the unscaled dispersive part of the response is much smaller for the low-concentration D condition than that for the C one. Here, in order to maximize resolution and clarity the y-axis-scale unit length is made greater than the x-axis one; so this is not quite a traditional complex-plane plot. Low-frequency electrode effects are somewhat less apparent for the present data than are those of Fig. 1, and the differences between the original-data curves and those representing only hopping arise primarily from the relative sizes of the dispersive contributions and those associated with the values of $\rho_0$ and $\varepsilon_{D\infty}$. As the limit of zero concentration of mobile ions is approached, simple non-dispersive Debye relaxation behavior stemming entirely from $\rho_0$ and $\varepsilon_{D\infty}$ becomes more and more dominant in the data, as discussed in Section II-B-2-a and demonstrated in detail below.

The second curve from the top of Fig. 2 shows scaled Debye response as a dashed line. Just below it appears the D-material points denoted "sub el" obtained after subtraction of electrode effects from the top D-data curve. The "sub el" points are exceptionally close indeed to the Debye response curve here. When the effect of $\varepsilon_{D\infty}$ is then subtracted from these points, however, the resulting D-material bulk dispersion points lie close to the U-model scaled response but show some scatter arising from stringent subtraction effects. Without such subtraction, however, the resulting original modulus formalism approach leads to a $\beta_1$ estimate of about 0.9 instead of 1/3 [6,7,20]. Because the C data set involves appreciably smaller electrode effects relative to dispersion effects than does the D data, its points after all subtractions lie somewhat closer to the U-model master curve than do the D ones, but their final dispersive hopping responses still show some scatter. Nevertheless, it is clear that both the C and D results scale to the U-model data curve.

Figure 3 presents scaled frequency-response results at the modulus level for the A and B situations. The scaled master curve and the subtracted points all lie appreciably above the original data points primarily because of the subtraction of the effects of $\varepsilon_{D\infty}$ [6,9]. The dispersive B points are poorer, however, than the A ones in the high-frequency region past the peak because electrode effects dominated the former data more than the latter, resulting in greater subtraction errors. Nevertheless, when the somewhat irregular B data points are fitted to the master curve with CNLS, the resulting open-circle points fit excellently.

Figure 4 shows similar results for the C and D material. Note that even with a magnification factor of 10 the original D data curve is appreciably smaller than the corresponding C one, resulting in greater deviations of the D dispersion points from the master curve than for the C points. Nevertheless, the results shown in Figs. 3 and 4 verify both the scaling approach and the appropriateness of the U fitting model.

In Fig. 5, a traditional log-log scaling plot involving scaled $\sigma'$ and scaled frequency is presented for all four fits included in Table I. In addition, a curve for a mixed alkali material with two types of mobile ions is included [35]. This curve was scaled to agree with the present master curve at its highest point, and it is evident that it then does not agree well with the single-ion U response curve, particularly in the low-frequency region where dispersion is just beginning to be evident.

The Fig.-5 plots of the scaled data and those in the magnified inset show appreciable electrode-polarization deviations from the master curve at low frequencies for all the data points, but note particularly the deviations appearing at high frequencies for the A data. The slope of this curve is increasing and reaches a value of about 0.77 at its highest point, in full agreement with prior work on non-negligible high-frequency electrode effects [8,36]. It is thus evident that electrode polarization can be important even at high frequencies where it may sometimes be erroneously identified as arising from nearly constant loss processes [8,36,37].

Finally, Fig. 6 shows scaled and fitted results for the dispersive-response parts of the A, B, C, and D data sets. Symbols of different sizes have been used to allow easy identification of the various responses. Although, as one would expect, the present scaling is limited only by the accuracy of the estimation of the scaling parameters from the original CNLS fits and is certainly near optimum, it is particularly gratifying that the estimated dispersive data points fit the master curve so well, thus verifying the appropriateness of the U model for these data sets. In the past, scaling has not usually been attempted for data that involve significant electrode polarization effects, but the present results show that this need not be a limitation, and, as well, it is clear that scaling is unnecessary except to allow comparison of plots for different situations. In most cases, one only need carry out CNLS fits of available data sets to obtain maximum information from them.

IV. SUMMARY

The present work shows that for homogeneous materials involving mobile ions of a single type the important $\beta_1$ shape parameter of the K1 dispersive frequency-response model has a unique, constant value of 1/3, resulting in the U model, one whose high-frequency-limiting log-log $\sigma'(\nu)$ slope is $n = 2/3$. These results are therefore inapplicable to mixed-alkali situations or to mixed electronic and ionic conduction. For single-ion materials, CNLS fits of frequency-response data with $\beta_1$ taken as a free parameter in the K1 model have led to estimates very close to 1/3. Here it is shown that on fixing $\beta_1$ at 1/3, the U model leads to excellent fits of data independent of temperature and ion-concentration variation, as expected from several different analyses [6-9]. Estimates of $n$ at high frequencies by others [23-25, 38-40] have led to $n \simeq 2/3$, independent of temperature and ionic concentration over the limited ranges considered, further confirmation of the appropriateness of the U model within its range of applicability.

Scaling, using the U-model fit results of Table I for variable concentration and temperature, was carried out for $\rho(\nu)$, $M''(\nu)$, and $\sigma'(\nu)$ data and resulted in the complex-plane and frequency-response plots of Figs. 1-6. Scaling was initially unsatisfactory for all these data sets because of the influence of non-dispersive effects associated with electrode polarization and with $\varepsilon_{D\infty}$. When these effects were subtracted to give best estimates of only the dispersive response, however, scaling of the resulting data was successful and led to data points close to those of the exact scaled U model. Fitting of these data points to this model then yielded scaled parameter values in excellent agreement with those of the model. Even when the best available scaling parameters are used, these results suggest that fitting with a model that takes all processes influencing the data into account may be necessary to yield meaningful scaling comparisons, as it certainly is in order to obtain good estimates of hopping and dielectric parameters from most data.

## ACKNOWLEDGMENTS

The author is most grateful to those who provided the data sets used in this study. The insightful comments and suggestions of Dr. J. C. Phillips and of a referee have been of great value.

Table I. Rows A-D: $\rho(\omega)$-level U-model CNLS fits to materials with different temperatures and ionic concentrations. The K0 line involves real-part K0 fitting, and MA stands for mixed alkali. All fits used modulus weighting except those of rows A and K0, where proportional weighting was used [19]. Here $100S_F$ is the percentage value of the relative standard deviation of a fit, and the last column lists its value for fits of the scaled and subtracted data to U-model hopping response..

| Type/ Ref. | Material | T (K) | $100S_F$ | $10^{-7}\rho_0$ ($\Omega$-cm) | $10^7\tau_o$ (s) | $\varepsilon_{C1\infty}$ | $\varepsilon_{D\infty}$ | $100\overline{S_F}$ or $[\beta_0]$ |
|---|---|---|---|---|---|---|---|---|
| | | | | | | | | |
| U | Scaled master: UM | ---- | ----- | $10^{-7}$ | $10^7$ | 6 | 0 | ----- |
| A/34 | 0.5Li•0.5La•TiO$_3$ | 179 | 0.98 | 6.25 | 233 | 25.27 | 65.01 | 1.65 |
| B/34 | 0.5Li•0.5La•TiO$_3$ | 225 | 0.42 | 0.018 | 0.467 | 17.72 | 80.13 | 1.29 |
| C/20 | 0.2K$_2$O•0.8GeO$_2$ | 414 | 1.13 | 22.8 | 87.4 | 2.60 | 9.29 | 1.68 |
| D/20 | 0.02K$_2$O•0.98GeO$_2$ | 602 | 0.45 | 24.8 | 2.82 | 0.077 | 9.51 | 1.99 |
| K0 | Fit $\sigma'(\overline{\nu})$ UM data | ---- | 6.6 | 9.6x10$^{-8}$ | 1.8x10$^8$ | 5.94 | ----- | [0.629] |
| MA/35 | 0.3(0.6Na$_2$O•0.4Li$_2$O) + 0.7B$_2$O$_3$ | Scaled to $\sigma'(\overline{\nu})$ UM data | | | | | | |

FIGURE CAPTIONS

1. Complex-plane resistivity plots of scaled data for the A (T= 179 K) and B (T=225 K) rows of Table I, before and after subtraction of all non-hopping contributions and including comparison of the latter results with exact U-model hopping response.

2. Stretched complex-plane resistivity plots of scaled variable-concentration data for the C and D rows of Table I, before and after subtraction of all non-hopping contributions for the C data (denoted "C sub") and separate subtractions of electrode effects ("sub el" points) and then of $\varepsilon_{D\infty}$ effects for the D data (denoted "D sub"). The sub el results are compared with pure Debye response, and the others with exact U-model dispersive response.

3. Scaled $\log_{10}$ frequency response of scaled $M''$ A and B data before and after subtraction of all non-hopping contributions and including comparison of the latter results with exact U-model hopping response. In addition, the open circles show the results of fitting the noisy subtracted B data with the U model.

4. Scaled $\log_{10}$ frequency response of scaled $M''$ C and D data before and after subtraction of all non-hopping contributions and including comparison of the latter results with exact U-model hopping response.

5. Log-log scaled frequency response of scaled $\sigma'$ A-D data sets, including comparison with exact U-model hopping response. The solid-circle points are for the mixed-alkali data identified in Table I. In addition, the low-frequency parts of the responses are shown with higher resolution in the inset graph.

6. Log-log scaled frequency response of scaled $\sigma'$ A, B, C, and D data sets fitted to the U-model master curve after subtraction of electrode-polarization contributions.

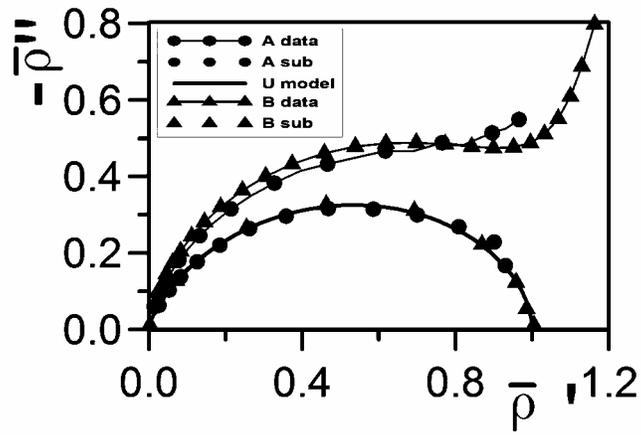

Fig.1

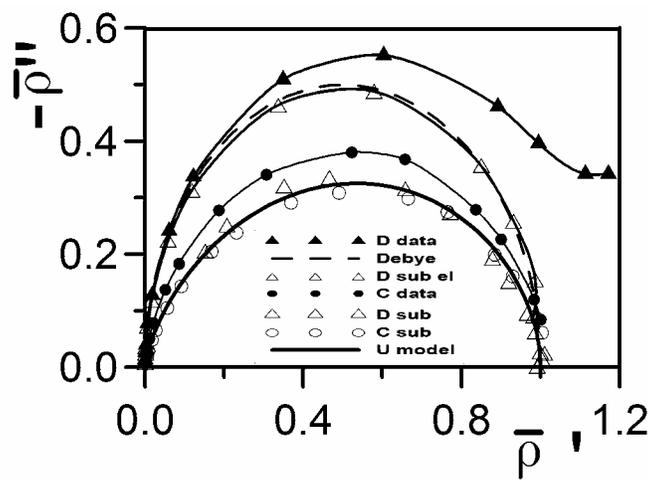

Fig.2

Fig. 3

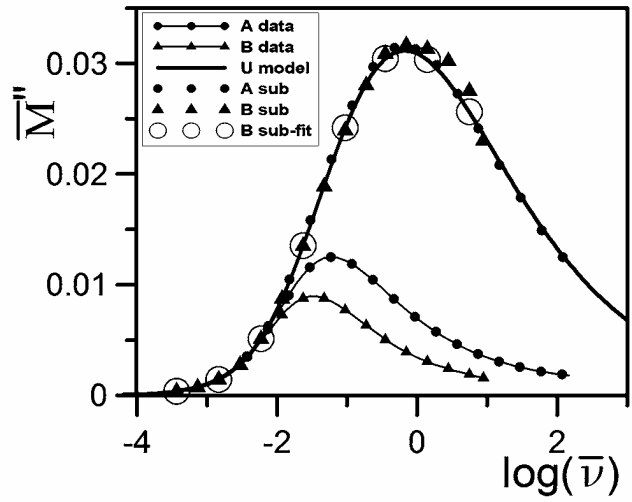

Fig. 4

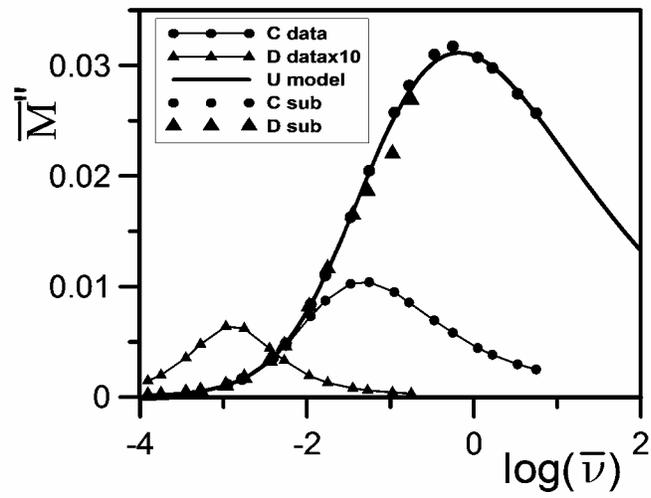

Fig. 5

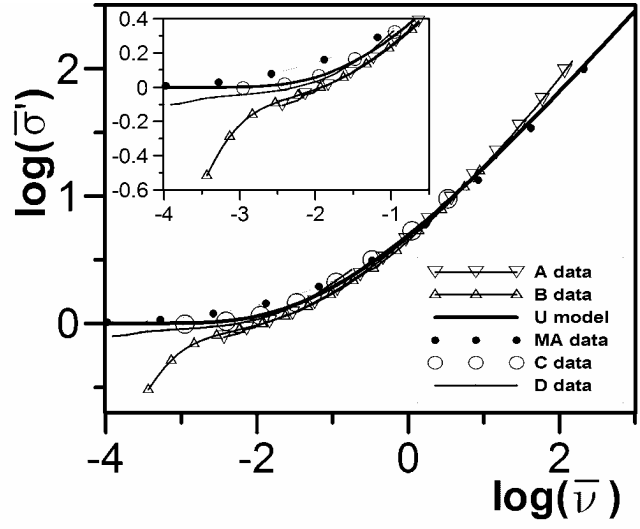

Fig. 6

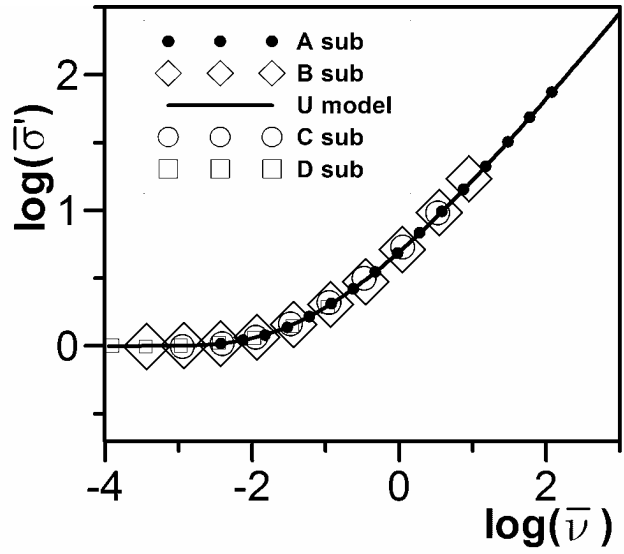